\documentclass[preprint]{elsarticle}

\usepackage{amssymb}
\usepackage{amsthm}

\usepackage{lineno}

\journal{Physica B}

\begin{document}

\begin{frontmatter}

\title{Bi-stability in single impurity Anderson model with strong electron-phonon interaction(polaron regime)}

\author[sbu]{Amir Eskandari-asl\corref{cor1}}
\ead{amir.eskandari.asl@gmail.com}
\address[sbu]{Department of physics, Shahid Beheshti University, G. C. Evin, Tehran 1983963113, Iran}

\begin{abstract}
We consider a single impurity Anderson model (SIAM) in which the quantum dot(QD) is strongly coupled to a phonon bath in polaron regime. This coupling results in an effective e-e attraction. By computing the self energies using a current conserving approximation which is up to second order in this effective attraction, we show that if the interaction is strong enough, in non particle-hole(PH) symmetric case, the system would be bi-stable and we have hysteresis loop in the I-V characteristic. Moreover, the system shows negative differential conductance in some bias voltage intervals. 
\end{abstract}

\begin{keyword}
SIAM \sep Polaron regime \sep bi-stability \sep negative differential conductance


\end{keyword}

\end{frontmatter}


\section{Introduction}
One of the important models to be studied in accordance with nano-electronics is SIAM which has been studied in many different aspects for many years\cite{Hewson}. One of the most important effects concerning this issue is the interaction of electrons with a phonon bath. The electron-phonon coupling in such small sized structures is usually noticeable and has to be considered in theoretical works. It is well known that such interactions result in the formation of polarons which attract each other, so effectively there is an attractive e-e interaction(see Ref.~\cite{Alexandrov1} and references there in).

One of the effects that is caused by the electron-phonon interaction in current carrying systems is the bi-stability. This issue has been under debate in recent years and there are several theoretical\cite{Gogolin1,Alexandrov1,Galperin1,Galperin2,Alexandrov2,Albrecht,Klatt} and experimental\cite{Elor,Lilj,CLi1} results confirming its existence. One of the open problems in this context is the existence of bi-stability in a system with a single level QD. In Ref.~\cite{Galperin2} , using Hartree-Fock approximation, it is shown that SIAM with electron-phonon coupling could show bi-stability. However, the authors of Ref.~\cite{Alexandrov2} claim that this appearing bi-stability is the artifact of the HF approximation. In the present work, we go beyond the HF approximation for our model by computing the self energies to the second order and look for the existence of bi-stability. We also see that our system shows negative differential conductance in some bias voltage intervals. 

Except for some simple cases as for example in Hartree approximation, the connection between the lesser Green function(GF) to the other GFs has not yet been determined and using the usual Langreth rules\cite{Haug} for analytical continuation will result in non conserved currents. Instead, as it is shown in Ref.~\cite{Taniguchi}, by demanding the current conservation, one could determine both the electron population and the current, by using the retarded GF only.

The paper is organized as follow: In Sec.2, after giving the Hamiltonian of the SIAM model, we describe our used approximation. In Sec.3 we present our numerical results for the non PH symmetric case and discuss them. Finally, Sec.4 concludes our work.  

\section{Hamiltonian model and the method}
  Our system consists of a single orbital QD connected to two leads. There exists an attractive e-e interaction which as already mentioned could stem from a strong coupling to a phonon bath\cite{Alexandrov1}. The Hamiltonian of this system reads:
\begin{eqnarray}
&&\hat{H}=\hat{H}_{dot}^{0}+\hat{H}_{leads}+\hat{H}_{tun}+\hat{H}_{e-e},
\label{eq:e1}
\end{eqnarray}
where
\begin{eqnarray}
&&\hat{H}_{dot}^{0}=\sum_{\sigma=\uparrow,\downarrow} eV_{g} \hat{c}_{\sigma}^{\dag} \hat{c}_{\sigma},
\label{eq:e2}
\\
&&\hat{H}_{leads}=\sum_{k\in\lbrace R,L\rbrace , \sigma=\uparrow,\downarrow}\epsilon_{k}\hat{a}_{k\sigma}^{\dag} \hat{a}_{k\sigma},
\label{eq:e3}
\\
&&\hat{H}_{tun}=\sum_{k\in\lbrace R,L\rbrace , \sigma=\uparrow,\downarrow} -t_{k} \hat{a}_{k\sigma}^{\dag} \hat{c}_{\sigma}+ h.c.,
\label{eq:e4}
\\
&&\hat{H}_{e-e}=-U_{0} \hat{n}_{\uparrow} \hat{n}_{\downarrow},
\label{eq:e5}
\end{eqnarray}
where $\hat{c}_{\sigma}(\hat{c}_{\sigma}^{\dag})$ annihilates(creates) an electron with spin $\sigma$ in the QD and $V_{g}$ is the gate voltage which determines the energy of the QD level. $\hat{a}_{k\sigma}(\hat{a}_{k\sigma}^{\dag})$ is the annihilation(creation) operator of an electron with spin $\sigma$ in the level $k$ of the leads, and $\epsilon_{k}$ is the energy of this level. $t_{k}$ is the spin independent hopping factor between the QD and the level k of the leads. $\hat{n}_{\sigma}$ is the number operator $\hat{c}_{\sigma}^{\dag} \hat{c}_{\sigma}$ and $U_{0}$ is the strength of the e-e interaction. We show the minus sign explicitly in the e-e interaction, so that for an attractive e-e interaction, $U_{0}$ would be positive.

The connections to the left and right leads are symmetric. Moreover, the leads are considered in WBL, so that
\begin{eqnarray}
\Gamma=2 \pi \sum_{k} \vert t_{k}\vert^{2} \delta(\epsilon-\epsilon_{k})
\label{eq:e6}
\end{eqnarray}
is independent of $\epsilon$. The bias voltage $V$ is applied symmetrically between the leads, so that $\mu_{L}=-\mu_{R}=eV/2$, where $\mu_{\alpha}$ $(\alpha=L,R)$ is the chemical potential of lead $\alpha$. We consider the system in zero temperature, so that the Fermi distribution of the leads are Heaviside theta functions: $\theta(\mu_{\alpha}-\epsilon)$. 

The Green function of the QD on the Keldysh time contour is defined as
\begin{eqnarray}
G_{\sigma}(\tau,\tau^{'})=-i \left\langle  T_{c} (\hat{c}_{\sigma}(\tau)\hat{c}_{\sigma}^{\dag}(\tau^{'})\right\rangle 
\label{eq:e7}
\end{eqnarray} 
where $\tau$ and $\tau^{'}$ lie on the Keldysh time contour and $ T_{c} $ is the contour ordering operator\cite{Zagoskin}. One could use the analytical continuation to obtain the standard retarded and advanced GFs and by Fourier transformation the corresponding GFs in the frequency domain would be found. Moreover, for a para magnetic solution, the GFs are spin independent.

The simplest way to consider the e-e interaction is the Hartree approximation (for our model there is no Fock term). In this approximation, the retarded, advanced, lesser and greater GFs, respectively, are
\begin{eqnarray}
&&G_{H\sigma}^{r}(\omega)=G_{H}^{a}(\omega)^{*}=\frac{1}{\omega-e V_{g}+n_{\sigma} U_{0}+i \Gamma /2},
\label{eq:e8}
\\
&&G_{H\sigma}^{<}(\omega)=\sum_{\alpha =R,L}  \frac{\left( i \Gamma /2\right)  \theta(\mu_{\alpha}-\omega)}{(\omega-e V_{g}+n_{\sigma} U_{0})^{2}+(\Gamma /2)^{2}},
\label{eq:e9}
\\
&&G_{H\sigma}^{>}(\omega)=G_{H\sigma}^{r}(\omega)-G_{H\sigma}^{a}(\omega)+G_{H\sigma}^{<}(\omega).
\label{eq:e10}
\end{eqnarray} 
If we wanted to use the self consistent Hartree approximation (as authors of Ref. \cite{Galperin1} did), we have to solve these equations self consistently with Eq.\ref{eq:e13}, but we improve the approximation to the second order. For the second order approximation, we use the Hartree GFs to construct a self energy and this self energy will be inserted into the Dyson equation to determine the final retarded and advanced GFs. On the Keldysh contour, the Dyson equation reads:
\begin{eqnarray}
G(\tau,\tau^{'})=G_{H}(\tau,\tau^{'})+\oint d\tau_{1} \oint d\tau_{2} G_{H}(\tau,\tau_{1}) \Sigma(\tau_{1},\tau_{2})G(\tau_{2},\tau^{'})\nonumber\\
\label{eq:e11}
\end{eqnarray}
where $ \Sigma(\tau_{1},\tau_{2}) $ is the self-energy.

 By analytical continuation using the Langreth rules\cite{Haug} and Fourier transformation we obtain the retarded and advanced GFs. For the second order approximation, the self energy contains terms up to the order of two in interaction and the GFs are
\begin{eqnarray}
&&G_{\sigma}^{r}(\omega)=G_{\sigma}^{a}(\omega)^{*}=\frac{1}{\omega-eV_{g}+n_{\sigma} U_{0}+i \Gamma /2-\Sigma_{\sigma}^{(2)r}(\omega)}. \nonumber\\
\label{eq:e12}
\end{eqnarray} 
For the half filled case where $ eV_{g}=U_{0}/2 $, $ n_{\sigma} $ is always equal to $ \frac{1}{2} $, so the terms with odd powers in interaction vanish in both Hartree and second order approximations. Therefore, up to these approximations the results for the repulsive and attractive interaction are the same for the PH symmetric case. However, for the non PH symmetric case the results for attractive interaction differ substantially.

\begin{figure}  
\includegraphics[width=8.5cm]{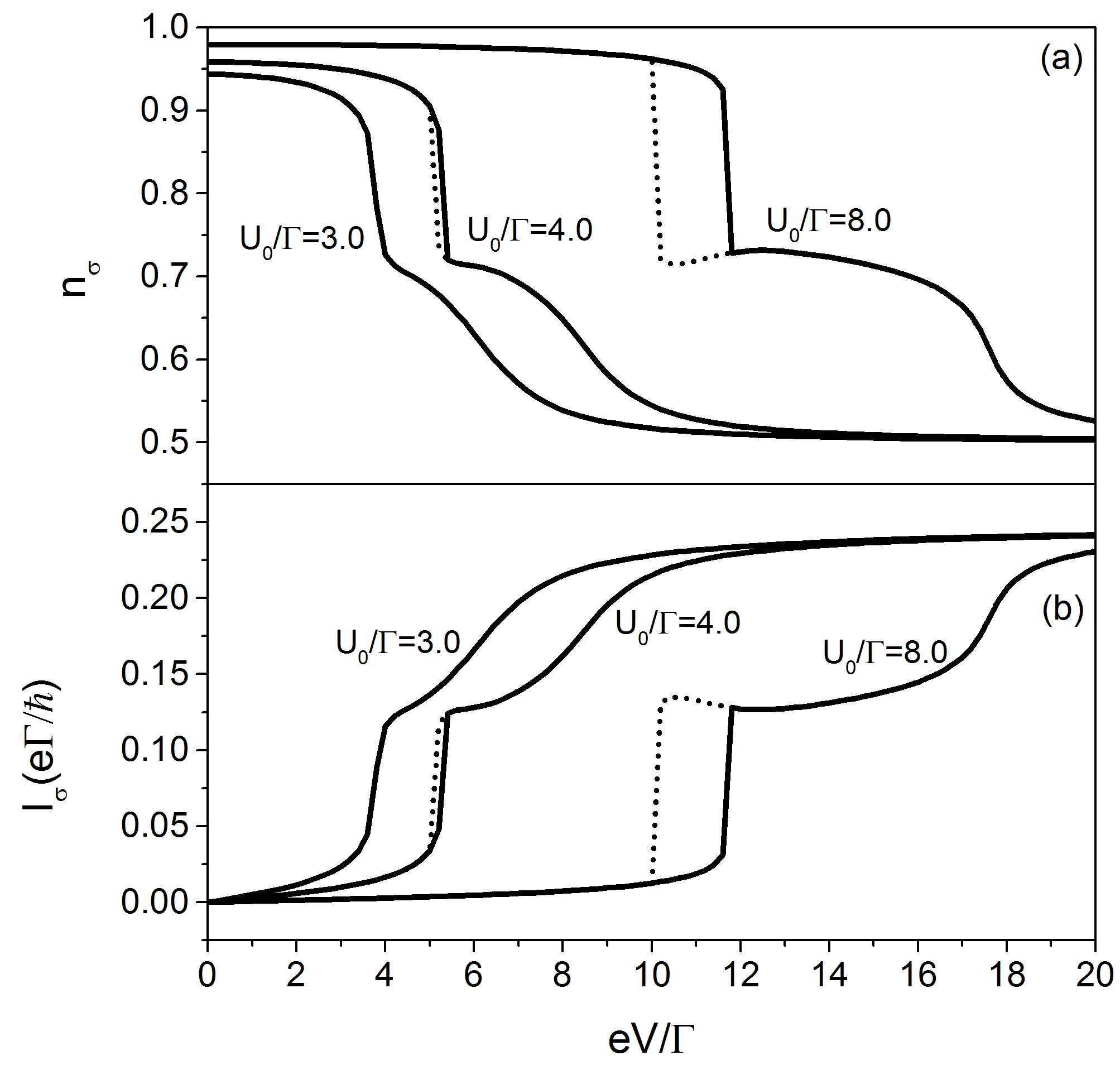}
\caption{\label{fig:one} (a)The electron populations and (b)the currents for the non PH symmetric case and $ U_{0}/\Gamma=3.0, 4.0, 8.0 $. The solid lines are in the forward direction and the dotted lines show the backward curves.}
\end{figure}  

As already mentioned, by demanding the current conservation, the formulas for the electron population and left to right current are obtained\cite{Taniguchi}. In our case, these formulas read:
\begin{eqnarray}
&&n_{\sigma}=- \int_{-\infty}^{\infty} \frac{d\omega}{2\pi} Im\left[ G^{r}_{\sigma}(\omega)\right] \sum_{\alpha=R,L}\theta(\mu_{\alpha}-\epsilon) ,
\label{eq:e13}
\\
&&I_{\sigma}=-\frac{e \Gamma}{2 \hbar}\int_{\mu_{R}}^{\mu_{L}} \frac{d\omega}{2 \pi} Im\left[ G^{r}_{\sigma}(\omega)\right]. 
\label{eq:e14}
\end{eqnarray}

For constructing the self energy we need the \textit{single-bubble} polarization which is
\begin{eqnarray}
\Pi_{\sigma}(\tau,\tau^{'})=-i G_{H\sigma}(\tau,\tau^{'}) G_{H\sigma}(\tau^{'},\tau).
\label{eq:e15}
\end{eqnarray}
Using Langreth rules for analytical continuation and doing Fourier transformation, the retarded, advanced and lesser polarizations could be numerically computed. For the WBL, as it is the case in our work, these polarizations could also be found analytically\cite{Taniguchi}.

The second order self energy on the Keldysh contour is 
\begin{eqnarray}
\Sigma_{\sigma}^{(2)}(\tau,\tau^{'})=i U_{0}^{2} \Pi_{\sigma}(\tau,\tau^{'}) G_{H\sigma}(\tau,\tau^{'})
\label{eq:e16}
\end{eqnarray} 
By analytical continuation and Fourier transformation, the retarded self energy reads:
\begin{eqnarray}
&&\Sigma_{\sigma}^{(2)r}(\omega)=\left( \Sigma_{\sigma}^{(2)a}(\omega) \right)^{*}=i U_{0}^{2} \int \frac{d\omega^{'}}{2 \pi} [\Pi_{\sigma}^{r}(\omega-\omega^{'}) G^{<}_{H\sigma}(\omega^{'})+\nonumber\\
&&\qquad \Pi_{\sigma}^{r}(\omega-\omega^{'}) G^{r}_{H\sigma}(\omega^{'})+\Pi_{\sigma}^{<}(\omega-\omega^{'}) G^{r}_{H\sigma}(\omega^{'})].
\label{eq:e17}
\end{eqnarray}
Substituting this self energy in Eq.\ref{eq:e12}, the second order GFs are obtained.

\section{Numerical Results}
In this section we present our results for the non PH symmetric(non-half filled) QDs only, since for the PH symmetric case, as already mentioned, the results are identical with those of the repulsive e-e interaction.

For simplicity, we take $ V_{g}=0 $. Using Eqs.\ref{eq:e12} and \ref{eq:e17}, we construct the retarded GF and with Eq.\ref{eq:e13}, find the electron population ($ n_{\sigma} $) self consistently. In Fig.\ref{fig:one}a we represent the electron populations and in Fig.\ref{fig:one}b the currents as functions of the applied bias voltages for $ U_{0}/\Gamma=3.0, 4.0, 8.0 $. For the cases where we have bi-stability and hysteresis loop, we use dotted as well as solid lines. Our numerical results suggest that in order to have bi-stability and consequently hysteresis loop in the I-V curve, the interaction strength has to be strong enough ($ U_{0}/\Gamma \gtrsim 3.5 $). Figs.\ref{fig:one}a and b show that by decreasing the population $ n_{\sigma} $ from near unity, the current increases. This can be understood by noting that when the QD is almost completely filled, its levels lie bellow the Fermi energies of the leads, so that the electrons could not tunnel from the QD to the right lead and the current is small. When the population reduces to near half, the QD levels lie between the Fermi energies of the leads and act as transport channels through which the electrons could make their way. 

\begin{figure}  
\includegraphics[width=8.5cm]{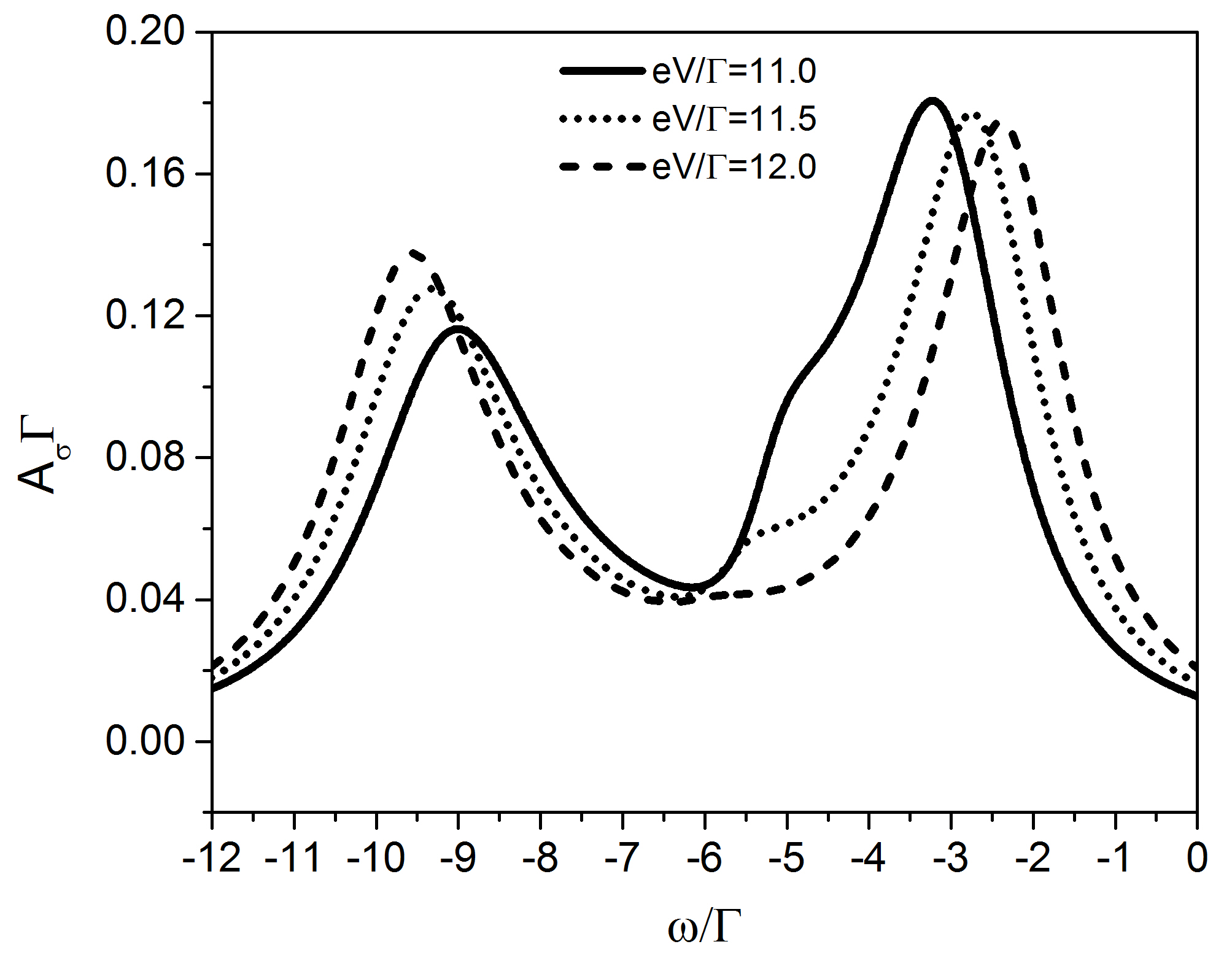}
\caption{\label{fig:two} The spectral function for  $ U_{0}/\Gamma=8.0 $ and $ eV/\Gamma=12.0,11.5,11.0 $ at the upper branch solution. As we see, by decreasing the bias voltage, a level near the Fermi energy of the right lead emerges, that is at the frequency of about $ \omega/\Gamma\approx-5.5 $.}
\end{figure}
 It is also seen in Fig.\ref{fig:one}b that by increasing the interaction strength, at some bias intervals, by decreasing the bias voltage, the current increases, that is the system shows negative differential conductance, which is the derivative of the current with respect to the bias voltage. For example, at $ U_{0}/\Gamma=8.0 $, the system shows negative differential conductance in the bias interval $ eV/\Gamma=10.5 $ to $ 12.4 $, at the upper branch solution. It is seen that by increasing the attraction strength, this negative differential conductance gets more significant. In order to understand this negative differential conductance, in Fig.\ref{fig:two}, we depict the spectral function ($ A_{\sigma}=-Im(G_{\sigma}^{r})/\pi $) as a function of the frequency for $ U_{0}/\Gamma=8.0 $ and $ eV/\Gamma=12.0,11.5,11.0 $ at the upper branch solution, that is when the electron population of one spin is close to 0.72. This spectral function plots show that in this region by decreasing the bias voltage, an effective level emerges at almost the Fermi energy of the right lead. This explains increasing current by decreasing bias voltage, since this emerging effective level acts as a transport channel.  
 
In order to clarify that these values of $ U_{0}/\Gamma $ are reachable in experiments, we note that the Haretree self energy, $ n_{\sigma} U_{0} $, (which appears in the denominator of Hartree GFs, Eq.\ref{eq:e8}), is actually the polaron shift of the QD level. Since $ n_{\sigma}$ is between 0.5 and 1, $ U_{0} $ is of the same order as the polaron shift (please note that for an electron-phonon coupling model like Refs.\cite{Galperin1} and \cite{Galperin3}, the polaron shift would be $ \frac{2 M^{2}}{\Omega} $ times electron population of QD, where $ M $ is the coupling strength and $ \Omega $ is phonon frequency). The observed values for polaron shift are 0.1-1 eV and values of $ \Gamma $ are also in this range (see Ref.\cite{Galperin3} and references there in. Please note that in their notation the polaron shift is of the order of $ E_{reorg} $). Therefor, the values of $ U_{0}/\Gamma $ that are used in our numerical computations could be achieved in experimental setups.

\section{Conclusions} 
In conclusion, we considered a single orbital QD with attractive e-e interaction that is connected to two leads. The attractive interaction stems from strong coupling to a phonon bath in polaron regime. By beginning from the Hartree GFs and constructing the second order self energy, we obtained the retarded and advanced GFs. Demanding current conservation, electron population and current could be obtained from retarded GF. By numerically solving the obtained relations, we plotted current and electron population as functions of the applied bias voltage. We saw that if the attraction is strong enough, the system shows bi-stability. On the other hand, there is a negative differential conductance which is explained by a level near the Fermi energy of the right lead (the drain). 

We didn't investigate the half filled case, since as we explained, for this case in the second order approximation the attractive interaction is identical to repulsive interaction which is discussed in Ref.~\cite{Taniguchi}    

\bibliographystyle{model1-num-names}
\bibliography{mp-005.bib}

\end{document}